\documentclass[11pt,a4paper]{article}
\usepackage[hyperref]{emnlp-ijcnlp-2019}

\usepackage[title]{appendix}

\usepackage{times}
\usepackage{url}
\usepackage{booktabs}
\usepackage{amsmath}
\usepackage{graphicx}
\usepackage{tikz}
\usepackage{pgfplots}

\aclfinalcopy

\definecolor{g-blue}{HTML}{2E86C1}
\definecolor{g-red}{HTML}{B03A2E}
\definecolor{g-purple}{HTML}{AF7AC5}

\title{Document Expansion by Query Prediction}

\author{Rodrigo Nogueira,$^1$ Wei Yang,$^2$ Jimmy Lin,$^2$ \and Kyunghyun Cho$^{3,4,5,6}$\vspace{0.1cm}\\
$^1$ Tandon School of Engineering, New York University \\
$^2$ David R. Cheriton School of Computer Science, University of Waterloo \\
$^3$ Courant Institute of Mathematical Sciences, New York University \\
$^4$ Center for Data Science, New York University \\
$^5$ Facebook AI Research~~
$^6$ CIFAR Azrieli Global Scholar \\
}

\begin{document}

\maketitle

\begin{abstract}
One technique to improve the retrieval effectiveness of a search engine is to expand documents with terms that are related or representative of the documents' content.
From the perspective of a question answering system, this might comprise questions the document can potentially answer.
Following this observation, we propose a simple method that predicts which queries will be issued for a given document and then expands it with those predictions with a vanilla sequence-to-sequence model, trained using datasets consisting of pairs of query and relevant documents.
By combining our method with a highly-effective re-ranking component, we achieve the state of the art in two retrieval tasks.
In a latency-critical regime, retrieval results alone (without re-ranking) approach the effectiveness of more computationally expensive neural re-rankers but are much faster.

Code to reproduce experiments and trained models can be found at \url{https://github.com/nyu-dl/dl4ir-doc2query}.
\end{abstract}

\section{Introduction}

The ``vocabulary mismatch'' problem, where users
use query terms that differ from those used in relevant documents, is one of the central challenges in information retrieval.
Prior to the advent of neural retrieval models, this problem has most often been tackled using query expansion techniques, where an initial round of retrieval can provide useful terms to augment the original query.
Continuous vector space representations and neural networks, however, no longer depend on discrete one-hot representations, and thus offer an exciting new approach to tackling this challenge.

\begin{figure}
\begin{center}
\centerline{\includegraphics[width=0.45\textwidth]{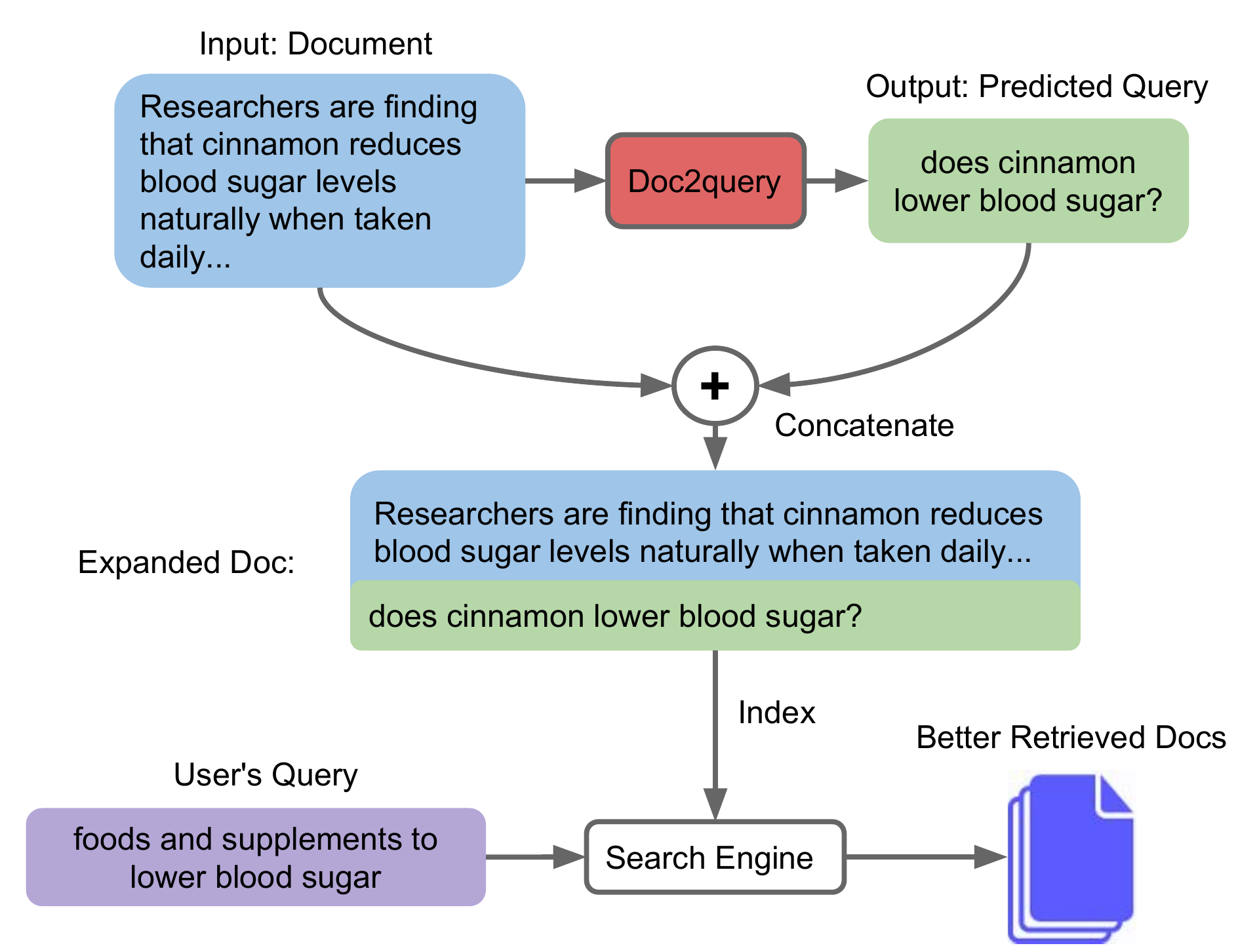}}
\vspace{-2mm}
\caption{Given a document, our Doc2query model predicts a query, which is appended to the document. Expansion is applied to all documents in the corpus, which are then indexed and searched as before.
}
\label{fig:overview}
\end{center}
\vspace{-10mm}
\end{figure}

Despite the potential of neural models to match documents at the semantic level for improved ranking, most scalable search engines use exact term match between queries and documents to perform initial retrieval.
Query expansion is about enriching the query representation while holding the document representation static.
In this paper, we explore an alternative approach based on enriching the document representation (prior to indexing).
Focusing on question answering, we train a sequence-to-sequence model, that given a document, generates possible questions that the document might answer.
An overview of the proposed method is shown in Figure~\ref{fig:overview}.

We view this work as having several contributions:\
This is the first successful application of document expansion using neural networks that we are aware of.
On the recent MS MARCO dataset~\cite{nguyen2016ms}, our approach is competitive with the best results on the official leaderboard, and we report the best-known results on TREC CAR~\cite{dietz2017trec}.
We further show that document expansion is more effective than query expansion on these two datasets.
We accomplish this with relatively simple models using existing open-source toolkits, which allows easy replication of our results.
Document expansion also presents another major advantage, since the enrichment is performed prior to indexing:\
Although retrieved output can be further re-ranked using a neural model to greatly enhance effectiveness, the output can also be returned as-is.
These results already yield a noticeable improvement in effectiveness over a ``bag of words'' baseline without the need to apply expensive and slow neural network inference at retrieval time.

\section{Related Work}

Prior to the advent of continuous vector space representations and neural ranking models, information retrieval techniques were mostly limited to keyword matching (i.e., ``one-hot'' representations).
Alternatives such as latent semantic indexing~\cite{LSI} and its various successors never really gained significant traction.
Approaches to tackling the vocabulary mismatch problem within these constraints include relevance feedback~\cite{Rocchio_1971}, query expansion~\cite{Voorhees:1994:QEU:188490.188508,Xu00}, and modeling term relationships using statistical translation~\cite{Berger:1999:IRS:312624.312681}.
These techniques share in their focus on enhancing {\it query} representations to better match documents.

In this work, we adopt the alternative approach of enriching {\it document} representations~\cite{tao2006language,pickens2010reverted,efron2012improving}, which works particularly well for speech~\cite{singhal1999document} and multi-lingual retrieval, where terms are noisy.
Document expansion techniques have been less popular with IR researchers because they are less amenable to rapid experimentation.
The corpus needs to be re-indexed every time the expansion technique changes (typically, a costly process); in contrast, manipulations to query representations can happen at retrieval time (and hence are much faster).
The success of document expansion has also been mixed; for example, \citet{billerbeck2005document} explore both query expansion and document expansion in the same framework and conclude that the former is consistently more effective.

A new generation of neural ranking models offer solutions to the vocabulary mismatch problem based on continuous word representations and the ability to learn highly non-linear models of relevance; see recent overviews by~\citet{Onal:2018:NIR:3229901.3229980} and~\citet{MitraBhaskar_Craswell_2019}.
However, due to the size of most corpora and the impracticality of applying inference over every document in response to a query, nearly all implementations today deploy neural networks as re-rankers over initial candidate sets retrieved using standard inverted indexes and a term-based ranking model such as BM25~\cite{robertson1995okapi}.
Our work fits into this broad approach, where we take advantage of neural networks to augment document representations prior to indexing; term-based retrieval then happens exactly as before.
Of course, retrieved results can still be re-ranked by a state-of-the-art neural model~\cite{nogueira2019passage}, but the output of term-based ranking already appears to be quite good.
In other words, our document expansion approach can leverage neural networks without their high inference-time costs.

\begin{table*}[t]
\centering\centering\resizebox{0.9\textwidth}{!}{
\begin{tabular}{l|ccc|c}
& TREC-CAR & \multicolumn{2}{c|}{MS MARCO} & Retrieval Time\\
& MAP & \multicolumn{2}{c|}{MRR@10} & ms/query\\
& Test & Test & Dev \\
\noalign{\vskip 1mm}
\hline
\noalign{\vskip 1mm}
Single Duet v2~\cite{mitra2019updated} & - & 24.5 & 24.3 &650$^\star$ \\
Co-PACRR$^{\spadesuit}$~\cite{macavaney2017contextualized} & 14.8 & - & - & - \\
\noalign{\vskip 1mm}
\hline
\noalign{\vskip 1mm}
BM25 & 15.3 & 18.6 & 18.4 & 50\\
BM25 + RM3 & 12.7 & - &16.7 & 250 \\
BM25 + Doc2query (Ours) & 18.3 & 21.8 & 21.5 & 90\\
BM25 + Doc2query + RM3 (Ours) &  15.5    & -  & 20.0 & 350\\

\noalign{\vskip 1mm}
\hline
\noalign{\vskip 1mm}
BM25 + BERT \cite{nogueira2019passage} & 34.8 & 35.9 & 36.5 & 3400$^{\dagger}$\\
BM25 + Doc2query + BERT (Ours) & \textbf{36.5} & \textbf{36.8} & \textbf{37.5} & 3500$^{\dagger}$\\
\end{tabular}}
\vspace{-2mm}
\caption{Main results on TREC-CAR and MS MARCO datasets. $^\star$ Our measurement, in which Duet v2 takes 600ms per query, and BM25 retrieval takes 50ms. $^{\spadesuit}$ Best submission of TREC-CAR 2017. $^{\dagger}$ We use Google's TPUs to re-rank with BERT. }
\label{tab:main_results}
\vspace{-2mm}
\end{table*}

\section{Method: Doc2query}

Our proposed method, which we call ``Doc2query'', proceeds as follows:\
For each document, the task is to predict a set of queries for which that document will be relevant.
Given a dataset of (query, relevant document) pairs, we use a sequence-to-sequence transformer model~\cite{vaswani2017attention} that takes as an input the document terms and produces a query.
The document and target query are segmented using BPE~\cite{sennrich2015neural} after being tokenized with the Moses tokenizer.\footnote{
\url{http://www.statmt.org/moses/}
} 
To avoid excessive memory usage, we truncate each document to 400 tokens and queries to 100 tokens.
Architecture and training details of our transformer model are described in Appendix~\ref{sec:architecture}.

Once the model is trained, we predict 10 queries using top-$k$ random sampling~\cite{fan2018hierarchical} and append them to each document in the corpus. 
We do not put any special markup to distinguish the original document text from the predicted queries. 
The expanded documents are indexed, and we retrieve a ranked list of documents for each query using BM25~\cite{robertson1995okapi}. 
We optionally re-rank these retrieved documents using BERT~\cite{devlin2018bert} as described by~\citet{nogueira2019passage}.

\section{Experimental Setup}

To train and evaluate the models, we use the following two datasets:

\smallskip \noindent {\bf MS MARCO}~\cite{nguyen2016ms} is a passage re-ranking dataset with 8.8M passages\footnote{
\url{https://github.com/dfcf93/MSMARCO/tree/master/Ranking}
} 
obtained from the top-10 results retrieved by the Bing search engine (from 1M queries).
The training set contains approximately 500k pairs of query and relevant documents.
Each query has one relevant passage, on average. The development and test sets contain approximately 6,900 queries each, but relevance labels are made public only for the development set. 

\smallskip \noindent {\bf TREC-CAR}~\cite{dietz2017trec} is a dataset where the input query is the concatenation of a Wikipedia article title with the title of one of its sections.
The ground-truth documents are the paragraphs within that section. 
The corpus consists of all English Wikipedia paragraphs except the abstracts.
The released dataset has five predefined folds, and we use
the first four as a training set (approx.\ 3M queries), and the remaining as a validation set (approx.\ 700k queries). 
The test set is the same as the one used to evaluate submissions to TREC-CAR 2017 (approx.\ 2,250 queries).

\smallskip \noindent We evaluate the following ranking methods:

\smallskip \noindent {\bf BM25}:\ We use the Anserini open-source IR toolkit~\cite{Yang:2017:AEU:3077136.3080721,Yang_etal_JDIQ2018}\footnote{
\url{http://anserini.io/}
} 
to index the original (non-expanded) documents and BM25 to rank the passages. During evaluation, we use the top-1000 re-ranked passages.

\smallskip \noindent {\bf BM25 + Doc2query}:\ We first expand the documents using the proposed Doc2query method. We then index and rank the expanded documents exactly as in the BM25 method above.

\smallskip \noindent {\bf RM3}:\ To compare document expansion with query expansion, we applied the RM3 query expansion technique~\cite{Abdul-Jaleel04}.
We apply query expansion to both unexpanded documents (BM25 + RM3) as well as the expanded documents (BM25 + Doc2query + RM3).

\smallskip \noindent {\bf BM25 + BERT}:\ We index and retrieve documents as in the BM25 condition and further re-rank the documents with BERT as described in \citet{nogueira2019passage}.

\smallskip \noindent {\bf BM25 + Doc2query + BERT}:\ We expand, index, and retrieve documents as in the BM25 + Doc2query condition and further re-rank the documents with BERT.


\smallskip \noindent To evaluate the effectiveness of the methods on MS MARCO, we use its official metric, mean reciprocal rank of the top-10 documents (MRR@10). For TREC-CAR, we use mean average precision (MAP).

\begin{table*}
\begin{center}
\begin{small}
\begin{tabular}{ll}
\noalign{\vskip 1mm}
\hline
\noalign{\vskip 1mm}
Input Document: & July is the hottest month in Washington DC with an average temperature of 27°C (80°F) and the coldest\\
& is January at 4°C (38°F) with the most daily sunshine hours at 9 in July. The wettest month is May with\\
& an average of 100mm of rain.\\
Predicted Query: & weather in washington dc\\
Target query: & what is the temperature in washington\\
\noalign{\vskip 1mm}
\hline
\noalign{\vskip 1mm}
Input Document: & The Delaware River flows through Philadelphia into the Delaware Bay. It flows through and aqueduct \\
& in the Roundout Reservoir and then flows through Philadelphia and New Jersey before emptying into\\
& the Delaware Bay.\\
Predicted Query: & what river flows through delaware\\
Target Query: & where does the delaware river start and end\\
\noalign{\vskip 1mm}
\hline
\noalign{\vskip 1mm}
Input Document: & sex chromosome - (genetics) a chromosome that determines the sex of an individual; mammals normally\\
& have two sex chromosomes chromosome - a threadlike strand of DNA in the cell nucleus that carries the\\
& genes in a linear order; humans have 22 chromosome pairs plus two sex chromosomes.\\
Predicted Query: & what is the relationship between genes and chromosomes\\
Target Query: & which chromosome controls sex characteristics\\
\noalign{\vskip 1mm}
\hline
\noalign{\vskip 1mm}
\end{tabular}
\end{small}
\end{center}
\vspace{-4mm}
\caption{Examples of query predictions on MS MARCO compared to real user queries.
}
\label{tab:examples}
\vspace{-2mm}
\end{table*}

\section{Results}

Results on both datasets are shown in Table~\ref{tab:main_results}.
BM25 is the baseline.
Document expansion with our method (BM25 + Doc2query) improves retrieval effectiveness by $\sim$15\% for both datasets.
When we combine document expansion with a state-of-the-art re-ranker (BM25 + Doc2query + BERT), we achieve the best-known results to date on TREC CAR; for MS MARCO, we are near the state of the art.\footnote{The top leaderboard entries do not come with system descriptions, and so it is not possible to compare our approach with theirs.}
Our full re-ranking condition (BM25 + Doc2query + BERT) beats BM25 + BERT alone, which verifies that the contribution of Doc2query is indeed orthogonal to that from post-indexing re-ranking.

Where exactly are these better scores coming from?
We show in Table~\ref{tab:examples} examples of queries produced by our Doc2query model trained on MS MARCO.
We notice that the model tends to copy some words from the input document (e.g., Washington DC, River, chromosome), meaning that it can effectively perform term re-weighting (i.e., increasing the importance of key terms).
Nevertheless, the model also produces words not present in the input document (e.g., weather, relationship), which can be characterized as expansion by synonyms and other related terms.

To quantify this analysis, we measured the proportion of predicted words that exist (copied) vs. not-exist (new) in the original document.
Excluding stop words, which corresponds to 51\% of the predicted query words, we found that 31\% are new while the rest (69\%) are copied.
If we expand MS MARCO documents using only new words and retrieve the development set queries with BM25, we obtain an MRR@10 of 18.8 (as opposed to 18.4 when indexing with original documents).
Expanding with copied words gives an MRR@10 of 19.7.
We achieve a higher MRR@10 of 21.5 when documents are expanded with both types of words, showing that they are complementary.

Further analyses show that one source of improvement comes from having more relevant documents for the re-ranker to consider.
We find that the Recall@1000 of the MS MARCO development set increased from 85.3 (BM25) to 89.3 (BM25 + Doc2query).
Results show that BERT is indeed able to identify these correct answers from the improved candidate pool and bring them to the top of the ranked list, thus improving the overall MRR.

As a contrastive condition, we find that query expansion with RM3 hurts in both datasets, whether applied to the unexpanded corpus (BM25 + RM3) or the expanded version (BM25 + Doc2query + RM3).
This is a somewhat surprising result because query expansion usually improves effectiveness in document retrieval, but this can likely be explained by the fact that both MS MARCO and CAR are precision oriented.
This result shows that document expansion can be more effective than query expansion, most likely because there are more signals to exploit as documents are much longer.

Finally, for production retrieval systems, latency is often an important factor.
Our method without a re-ranker (BM25 + Doc2query) adds a small latency increase over baseline BM25 (50 ms vs.\ 90 ms) but is approximately seven times faster than a neural re-ranker that has a three points higher MRR@10 (Single Duet v2, which is presented as a baseline in MS MARCO by the organizers).
For certain operating scenarios, this tradeoff in quality for speed might be worthwhile.

\section{Conclusion}

We present the first successful use of document expansion based on neural networks.
Document expansion holds substantial promise for neural models because documents are much longer and thus contain richer input signals.
Furthermore, the general approach allows developers to shift the computational costs of neural network inference from retrieval to indexing.

Our implementation is based on integrating three open-source toolkits:\ OpenNMT~\cite{opennmt}, Anserini, and TensorFlow BERT.
The relative simplicity of our approach aids in the reproducibility of our results and paves the way for further improvements in document expansion.

\section*{Acknowledgments}

KC thanks support by NVIDIA and CIFAR and was partly supported by Samsung Advanced Institute of Technology (Next Generation Deep Learning: from pattern recognition to AI) and Samsung Electronics (Improving Deep Learning using Latent Structure).
JL thanks support by the Natural Sciences and Engineering Research Council (NSERC) of Canada.

\bibliographystyle{ACM-Reference-Format}
\bibliography{main}

\clearpage
\begin{appendices}

\section{Architecture and Training Details}
\label{sec:architecture}

The architecture of our transformer model is identical to the \textit{base} model described in \citet{vaswani2017attention}, which has 6 layers for both encoder and decoder, 512 hidden units in each layer, 8 attention heads and 2048 hidden units in the feed-forward layers.
We train with a batch size of 4096 tokens for a maximum of 30 epochs. 
We use Adam~\cite{kingma2014adam} with a learning rate of $10^{-3}$, $\beta_1$ = 0.9, $\beta_2$ = 0.998, L2 weight decay of 0.01, learning rate warmup over the first 8,000 steps, and linear decay of the learning rate. We use a dropout probability of 0.1 in all layers. 
Our implementation uses the OpenNMT framework~\cite{opennmt}; training takes place on four V100 GPUs.
To avoid overfitting, we monitor the BLEU scores of the training and development sets and stop training when their difference is larger than four points.

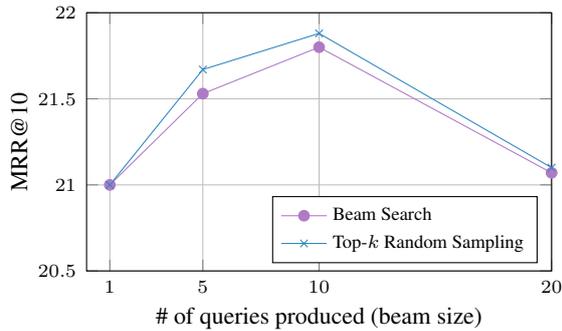
\begin{figure}[t]
\centering
\begin{tikzpicture}
\begin{axis}[
width=1.0\columnwidth,
height=0.65\columnwidth,
legend cell align=left,
legend style={at={(1, -0.1)},anchor=south east,font=\scriptsize},
mark options={mark size=3},
font=\scriptsize,
xmin=0, xmax=20,
ymin=20.5, ymax=22,
xtick={1, 5, 10, 20},
ytick={20.5, 21.0, 21.5, 22},
legend pos=south east,
xmajorgrids=true,
ymajorgrids=true,
xlabel style={font = \small, yshift=1ex},
xlabel=\# of queries produced (beam size),
ylabel=MRR@10,
ylabel style={font = \small, yshift=-2ex}
]    legend entries={Recall,
        Top 1 EM},
    ]
    \addplot[mark=*,g-purple, mark options={scale=1}] plot coordinates {
   (1, 21)(5, 21.53)(10, 21.8)(20, 21.07)
    };
    \addlegendentry{Beam Search}
	\addplot[mark=x,g-blue, mark options={scale=1}] plot coordinates {
	   (1, 21)(5, 21.67)(10, 21.88)(20, 21.1)
	    };
    \addlegendentry{Top-$k$ Random Sampling}
    \end{axis}
    
    \end{tikzpicture}
\caption{Retrieval effectiveness on the development set of MS MARCO when using different decoding methods to produce queries. On the {\it x}-axis, we vary the number of predicted queries that are appended to the original documents.} 
\label{fig:decoding_comparison}
\vspace{0mm}
\end{figure}

\section{Evaluating Various Decoding Schemes}

Here we investigate how different decoding schemes used to produce queries affect the retrieval effectiveness.
We experiment with two decoding methods:\ beam search and top-$k$ random sampling with different beam sizes (number of generated hypotheses). Results are shown in Figure~\ref{fig:decoding_comparison}.
Top-$k$ random sampling is slightly better than beam search across all beam sizes, and we observed a peak in the retrieval effectiveness when 10 queries are appended to the document.
We conjecture that this peak occurs because too few queries yield insufficient diversity (fewer semantic matches) while too many queries introduce noise and reduce the contributions of the original text to the document representation.

\end{appendices}

\end{document}